\documentclass[a4paper]{article}

\usepackage{INTERSPEECH2021}

\usepackage{pifont}
\newcommand{\cmark}{\ding{51}}%
\newcommand{\xmark}{\ding{55}}%

\usepackage{booktabs}
\usepackage{tabularx}
\usepackage{multirow}
\newcommand{\mytable}{
	\centering
	\renewcommand{\arraystretch}{1.1}
}

\newcolumntype{C}{>{\centering\arraybackslash}X}
\newcolumntype{L}{>{\raggedright\arraybackslash}X}
\newcolumntype{R}{>{\raggedleft\arraybackslash}X}
\newcolumntype{P}[1]{>{\raggedright\arraybackslash}p{#1}}

\newcommand{\PreserveBackslash}[1]{\let\temp=\\#1\let\\=\temp}
\newcolumntype{A}[1]{>{\PreserveBackslash\raggedright}p{#1}}
\newcolumntype{B}[1]{>{\PreserveBackslash\centering}p{#1}}

\usepackage{microtype}
\usepackage{url}

\usepackage{cite}

\title{Analyzing Speaker Information in Self-Supervised Models to Improve Zero-Resource Speech Processing}
\name{Benjamin van Niekerk, Leanne Nortje, Matthew Baas, Herman Kamper}
\address{E\&E Engineering, Stellenbosch University}
\email{\footnotesize \{benjamin.l.van.niekerk,nortjeleanne,matthew.baas\}@gmail.com, kamperh@sun.ac.za}

\begin{document}

\maketitle
\begin{abstract}
Contrastive predictive coding (CPC) aims to learn representations of speech by distinguishing future observations from a set of negative examples.
Previous work has shown that linear classifiers trained on CPC features can accurately predict speaker and phone labels. 
However, it is unclear how the features actually capture speaker and phonetic information, and whether it is possible to normalize out the irrelevant details (depending on the downstream task). 
In this paper, we first show that the per-utterance mean of CPC features captures speaker information to a large extent. 
Concretely, we find that comparing means performs well on a speaker verification task. 
Next, probing experiments show that standardizing the features effectively removes speaker information. 
Based on this observation, we propose a speaker normalization step to improve acoustic unit discovery using $K$-means clustering of CPC features. 
Finally, we show that a language model trained on the resulting units achieves some of the best results in the ZeroSpeech2021~Challenge.
\end{abstract}
\noindent\textbf{Index Terms}: unsupervised speech processing, self-supervised learning, acoustic unit discovery, spoken language modeling.

\section{Introduction}

{A core goal of \textit{zero-resource speech processing} is to develop methods that can learn robust representations of speech without supervision~\cite{jansen+etal_icassp13, dunbar+etal_asru17, dunbar+etal_interspeech19, dunbar+etal_interspeech20}.}
{These representations can be used to bootstrap training in downstream speech systems and reduce requirements on labeled data~\cite{kahn_icassp20, chung+etal_interspeech19, baevski+etal_neurips20}.}
{While a range of self-supervised methods have been developed for speech~\cite{heck+etal_asru17, chung+etal_interspeech19, baevski+etal_neurips20, baevski+etal_iclr20, chorowski+etal_taslp19, wang+etal_icassp20}, in this paper we focus on contrastive predictive coding (CPC)~\cite{vandenoord+etal_arxiv18}}.

CPC models are trained to distinguish future observations from a set of negative examples.
{The idea is that to accurately identify future speech segments, the model must learn meaningful phonetic contrasts while being invariant to low-level details such as background noise.}
{Recent studies~\cite{vandenoord+etal_arxiv18, vanniekerk+etal_interspeech20} show that separate linear classifiers trained on CPC features can accurately predict both speaker and phone categories.}
Features that capture either phonetic or speaker information can be useful depending on the downstream task.
{However, it is unclear whether we can disentangle or discard either component from the CPC features e.g., if speaker-invariance is required for a specific task.}

In this paper, we investigate how speaker information is represented in CPC features.
We qualitatively (Section~\ref{sec:cpc_analysis}) and quantitatively (Section~\ref{sec:speaker_verification}) show that the per-utterance mean over CPC features captures a large degree of the speaker information.
Based on this observation, we propose a simple speaker normalization step that effectively removes speaker information (Section~\ref{sec:probing}).
We then show that speaker normalization improves performance on two downstream tasks: acoustic unit discovery~\cite{varadarajan+etal_acl08, lee+glass_acl12, ondel+etal_pcs16, vanniekerk+etal_interspeech20}, and spoken language modeling~\cite{nguyen_zero_2020, lakhotia_etal_arxiv21}.
Specifically, we improve an acoustic unit discovery system based on $K$-means clustering of CPC features (Section~\ref{sec:aud}) and show that an LSTM-based language model trained on the discovered units achieves some of the best scores in the ZeroSpeech2021 challenge~\cite{dunbar+etal_interspeech21} (Section~\ref{sec:lm}).

{Our speaker normalization approach is very simple, making it easy to incorporate into current and future CPC speech models.}

\begin{figure*}[t]
\includegraphics[width=\textwidth]{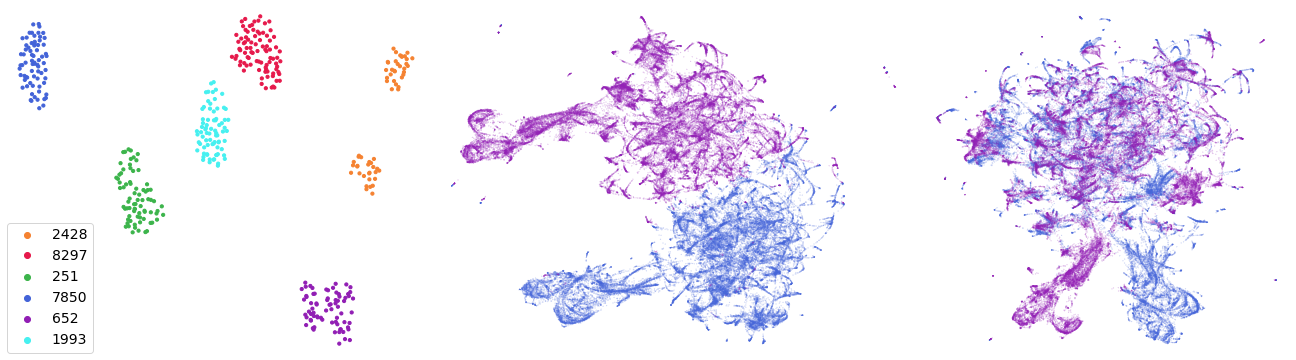}
\caption{UMAP visualizations of CPC features. (a) The per-utterance means of CPC features for six speakers. (b) Per-frame CPC features for the blue and purple speakersin (a). (c) Per-frame CPC features (standardized per utterance) for the same speakers.}
\label{fig:umap}
\vspace{-4mm}
\end{figure*}

\section{Analysis of CPC features}
\label{sec:cpc_analysis}

\subsection{Contrastive predictive coding}
CPC models consist of two components: an encoder, and a context network.
First, the encoder maps input audio into a sequence of embeddings $(z_1, \ldots, z_T)$.
Next, the autoregressive context network summarizes the embeddings (up until time $t$) into a context vector $c_t$.
Using this context, the model is trained to discriminate actual future embeddings from a set of negative examples drawn from other utterances.
Specifically, we minimize the constrastive loss:
\[
\mathcal{L}_t := -\frac{1}{M} \sum_{m=1}^M \log \left[\frac{\exp(z_{t+m}^\intercal W_m c_t)}{\sum_{\tilde{z} \in \mathcal{N}_{t, m}} \exp(\tilde{z}^\intercal W_m c_t)} \right],
\]
where $M$ is the prediction horizon, $W_m$ is a linear classifier, and $\mathcal{N}_{t, m}$ is a set containing the negative examples along with the correct future embedding $z_{t+m}$.

In this paper, we use the CPC-big model from \cite{nguyen_zero_2020} trained on the LibriLight \texttt{unlab-6k} set~\cite{kahn_icassp20}. 
The encoder consists of five
convolutional layers each with $512$ channels, kernel sizes $\langle 10, 8, 4, 4, 4 \rangle$, and strides $\langle 5,4,2,2,2 \rangle$.
Given raw audio sampled at $16$ kHz, the encoder extracts embeddings with a hop length of 10 ms.
The context network is a stack of four LSTM layers with $512$ hidden units each.
Finally, the linear classifier $W_m$ is replaced with a single-layer %1-layer
transformer.
We use the outputs of the second LSTM layer as speech features since they gave the best ABX phone discrimination results in~\cite{nguyen_zero_2020}.
In the remainder of the paper we refer to these as the \textit{CPC features}.

\subsection{A visual exploration of the CPC features}

Previous work~\cite{vandenoord+etal_arxiv18, vanniekerk+etal_interspeech20} has shown that CPC features capture both phonetic and speaker information.
However, it is unclear how the representation structures this information.
We hypothesize that the per-utterance mean of the features captures a large degree of the speaker information.
This is reasonable under the assumption that speaker identity remains constant over an utterance with phonetic content varying over shorter time scales~\cite{hsu+etal_interspeech18}.

As a first step towards validating this hypothesis, we explore the CPC features using UMAP~\cite{mcinnes+etal_joss18}.
Figure~\ref{fig:umap}(a) shows the per-utterance mean of CPC features for six speakers selected from the LibriSpeech \texttt{dev-clean} set~\cite{panayotov_etal+icassp15}.
The different speakers are clearly separated, showing that the mean does indeed capture speaker information.
In Figure~\ref{fig:umap}(b) we zoom in, visualizing the CPC features as individual frames for two speakers (colored blue and purple).
Although the UMAP embeddings for the two speakers exhibit similar structure, they are still separated based on speaker identity. 
This contrasts with Figure~\ref{fig:umap}(c) which shows the same features after standardization, i.e.\ per-utterance mean and variance normalization of the CPC features.
Here the structures are more aligned and no longer separated by speaker.

\section{Speaker verification}
\label{sec:speaker_verification}

In this section, we verify quantitatively that the per-utterance means of the CPC features capture speaker information.
We show that simply comparing the means performs well on a speaker verification task.
Given a set of enrollment utterances, the goal of speaker verification is to determine whether a new utterance belongs to a specific speaker.
To set up the task, we randomly select five enrollment utterances for each speaker in the LibriSpeech \texttt{dev-clean} set, reserving the remainder for testing.
We compare three systems across two metrics: classification accuracy and equal error rate (EER). 

The first system is based on the means of the CPC features.
In the enrollment step, we extract CPC features and compute the mean for each utterance.
The means are then aggregated to find a single speaker embedding.
At test time, we use Euclidean distance to compare the CPC feature mean of an utterance to the reference speaker embeddings.
For classification accuracy, we select the closest speaker as the prediction.
For EER, we threshold the distance to decide if the test utterance matches a given speaker.

The second system is a naive baseline that follows the same approach, but uses Mel-frequency cepstral coefficients (MFFCs) instead of CPC features.
This system should provide a lower bound on the performance of the CPC-based approach.

The third system is a supervised topline based on the GE2E loss~\cite{wan_generalized_2020}.
We use an open-source implementation trained on more than 8k speakers.\footnote{\scriptsize \url{https://github.com/resemble-ai/Resemblyzer}}
This system was specifically trained for speaker verification using a discriminative loss on a much larger dataset.
Therefore it serves as an upper bound on expected performance.

Table~\ref{tbl:verification} shows the results for the three approaches.
The CPC-based system clearly outperforms the baseline.
While there is a gap in performance compared to the topline, our goal was to demonstrate that the per-utterance mean of CPC features results in discriminative speaker embeddings.

\begin{table}[!t]
	\vspace{1mm}
	\mytable
	\caption{Speaker verification results for the supervised topline and the CPC- and MFCC-based systems.}
	\eightpt % Interspeech
	\vspace{-2mm}
	\begin{tabularx}{1.0\linewidth}{@{}lCC@{}}
		\toprule
		& EER (\%) & Accuracy (\%) \\
		\midrule
		Topline: GE2E & \hphantom{0}1.6 & 98.8 \\
		Proposed: Mean of CPC & \hphantom{0}6.7 & 95.8 \\
		Baseline: Mean of MFCCs & 19.8 & 59.8 \\
		\bottomrule
	\end{tabularx}
	\label{tbl:verification}
	\vspace{-5mm}
\end{table}

\section{Speaker normalization}

Based on the above observations, we propose standardizing  the CPC features as a simple speaker normalization step.
Given an utterance (or set of utterances) from a single speaker, we remove speaker information from the CPC features by subtracting the mean and scaling to unit variance. 
In the remainder of the paper, we analyze this speaker normalization step and apply it to two downstream tasks: acoustic unit discovery, and spoken language modeling (see Figure~\ref{fig:approach_task}).
\vspace{-1mm}

\subsection{Acoustic unit discovery}

In contrast to continuous representation learning, acoustic unit discovery involves finding a set of discrete units corresponding to the phonetic inventory of a language~\cite{dunbar+etal_interspeech20, lee+glass_acl12}.
We incorporate speaker normalization into a baseline acoustic unit discovery system built on $K$-means clustering applied to CPC features~\cite{nguyen_zero_2020}.
Concretely, we cluster the speaker-normalized features using $K$-means with 50 clusters.
The cluster means are estimated on speech from a subset of 35 speakers in the LibriSpeech \texttt{train-clean-100} set.
\vspace{-1mm}

\subsection{Spoken language modeling}

\begin{figure}[!t]
    \centering
    \includegraphics[scale=0.8]{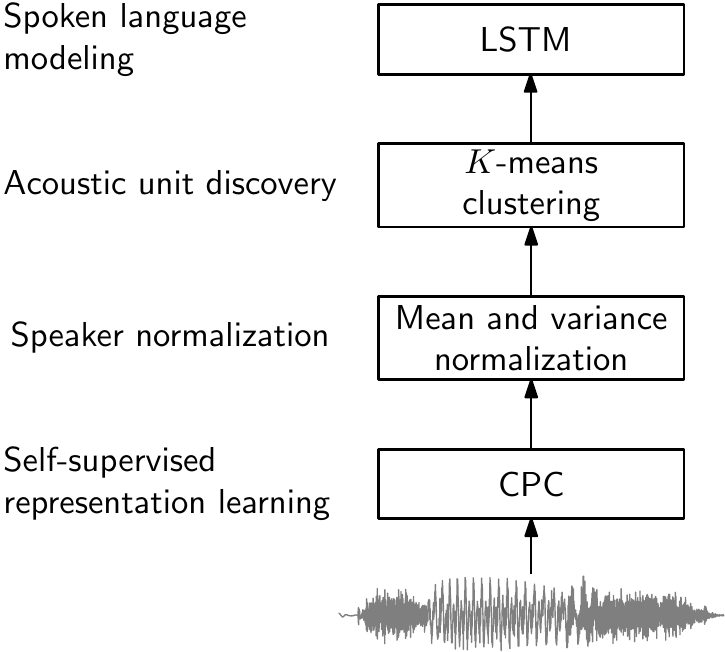}
    \caption{We propose a speaker normalization method for CPC features. We incorporate speaker normalization into an acoustic unit discovery system (based on $K$-means clustering) and spoken language model (trained on the clustered codes).}
    \label{fig:approach_task}
    \vspace{-5mm}
\end{figure}

To determine whether the discovered acoustic units capture structure beyond just the acoustics, 
we consider the task of spoken language modeling~\cite{nguyen_zero_2020, lakhotia_etal_arxiv21}.
For this task, we train an LSTM language model on the units (top of Figure~\ref{fig:approach_task}).
We use the language model architecture of~\cite{nguyen_zero_2020}:
three LSTM layers each with 1024 hidden units.
We train for 100k steps with a batch size of 32k acoustic unit tokens.
To evaluate the quality of the language model, we use several metrics developed specifically for the goal of language modeling on speech (Section~\ref{sec:lm}).
Most of these metrics rely on a score for how probable a spoken segment is under the language model.
To score a spoken segment, we first encode it into a sequence of acoustic units $(q_1, \ldots, q_N)$, based on the trained $K$-means model.
We then compute the log probability of the sequence using the chain rule, $\log P(q_1, \ldots, q_N) = \sum_{k = 1}^N \log P(q_k | q_1, \ldots, q_{k-1})$,
where $P(q_k | q_1, \ldots, q_{k-1})$ is the output of the LSTM.

\section{Probing experiments}
\label{sec:probing}

\begin{table}[!b]
	\vspace{-4mm}
    \mytable
    \caption{Probing experiments where phone, speaker and gender classifiers are trained on CPC features. Clustering is performed on the CPC features using $K$-means with 50 clusters.}
    \eightpt % Interspeech
    \begin{tabularx}{1.0\linewidth}{@{}CCCCC@{}}
        \toprule
        & & \multicolumn{3}{c}{Accuracy (\%)} \\
        \cmidrule(l){3-5}
        Standardized & Clustered & Phone & Speaker & Gender \\
        \midrule
        \multicolumn{2}{@{}l}{\textit{{Linear classifiers:}}} \\[2pt]
        % \multicolumn{4}{l}{\underline{\textit{Linear probe}}} \\
        \xmark & \xmark & 75.7 & 93.4 & 96.7 \\
        \cmark & \xmark & 77.0 & 14.8 & 55.3 \\
        \xmark & \cmark & 46.6 & \hphantom{0}3.4 & 53.5 \\
        \cmark & \cmark & 48.5 & \hphantom{0}3.1 & 50.9 \\[5pt]
        \midrule
        \multicolumn{2}{@{}l}{\textit{{Non-linear classifiers:}}} \\[2pt]        
        \xmark & \xmark & 80.1 & 99.5 & 99.8 \\
        \cmark & \xmark & 79.7 & 89.0 & 98.1 \\
        \bottomrule
    \end{tabularx}
    \label{tbl:probe}
\end{table}

{To evaluate the speaker normalization step and further analyze the information captured by the CPC features, we conduct a series of probing experiments.}
We train a set of classifiers to predict either phone, speaker, or gender labels given a single CPC feature frame as input.
The classifiers are trained on the \texttt{dev-clean} subset of LibriSpeech with classification accuracy reported on a held-out set of 400 utterances (10 per speaker).
For the phone classifier, we use the Montreal Forced Aligner~\cite{mcauliffe_montreal_2017} to extract time-aligned phone labels (from 41 phone classes).

We first train a set of \textit{linear} classifiers. The results (averaged over 10 runs) are shown at the top part of Table~\ref{tbl:probe}.
These classifiers assess the degree to which phone, speaker, and gender classes are linearly separable.
In the first row we repeat the experiments in~\cite{vandenoord+etal_arxiv18,vanniekerk+etal_interspeech20}, showing that CPC features linearly separate phonetic, gender and speaker information.
The second row shows that standardizing the features degrades speaker and gender classification accuracy, but slightly improves phone classification.
This is reasonable given our observation that the mean captures speaker identity.

How does standardization affect results for \textit{clustered} CPC features?
This question is of interest for acoustic unit discovery (see Section~\ref{sec:aud}).
In the third and fourth rows of Table~\ref{tbl:probe}, we repeat the linear classifier experiments but on clustered features (using $K$-means with 50 clusters).
We see that the clustering step reduces accuracy across all tasks.
In particular, speaker and gender accuracy is reduced almost to the level of chance ($2.5\%$ and $50\%$ respectively).
This shows that the clusters primarily capture phonetic information with standardization improving performance.

The experiments above investigated linear separability, but to what extent do the features capture phone, speaker and gender properties more generally?
To answer this question, we train non-linear classifiers on the CPC features.
Specifically, we use multi-layer perceptrons with one hidden ReLU-layer containing 1024 units.
Results are shown at the bottom of Table~\ref{tbl:probe}.
Overall the non-linear classification scores are higher than their linear counterparts.
However, standardization still reduces speaker classification accuracy (by over 10\% absolute). 
On the other hand, phone and gender accuracies remain similar.
This indicates that while standardization removes speaker and gender information in the linear case, these characteristics are still present in the features (albeit non-linearly).

\section{Results on downstream tasks}
\label{sec:aud}

\subsection{Acoustic unit discovery}

\textbf{ABX phone discrimination tests.} 
In this section, we use ABX phone discrimination tests~\cite{schatz+etal_interspeech13} to evaluate the acoustic unit discovery system.
These tests ask whether triphone $X$ is more similar to triphone $A$ or $B$. 
Here $A$ and $X$ are instances of the same triphone (e.g.\ ``beg''), while $B$ differs in the middle phone (e.g.\ ``bag'').
For the \textit{within-speaker} test, $A$, $B$, and $X$, are all taken from the same speaker.
The \textit{across-speaker} test aims to measure speaker-invariance by taking $A$ and $B$ from the same speaker, but $X$ from a different speaker.  
ABX is reported as an aggregated error rate over pairs of triphones.
For the similarity metric between encoded segments, we use the average cosine distance along the dynamic time warping alignment path.

Table~\ref{tbl:abx} shows ABX results on the \texttt{dev-clean} and \texttt{dev-other} subsets of LibriSpeech.
Without clustering (rows one and two), speaker normalization slightly improves ABX scores.
For the clustered CPC features (rows three and four), ABX is performed over one-hot encoded cluster codes.
In this case, speaker normalization improves ABX by more than 13\% relative on both the within and across speaker tests.

\begin{table}[!b]
	\vspace{-4mm}
	\mytable
	\caption{ABX error rates for CPC features and MFCCs.}
	\eightpt
	\begin{tabularx}{1.0\linewidth}{@{}CCcccc@{}}
		\toprule
		& & \multicolumn{2}{c}{Within (\%)} & \multicolumn{2}{c}{Across (\%)} \\
%		\cline{3-4} \cline{5-6}
        \cmidrule(l){3-4} \cmidrule(l){5-6}
		Standardized & Clustered & clean & other & clean & other \\
		\midrule
%        \midrule
        \multicolumn{2}{@{}l}{\textit{{CPC features:}}} \\[2pt]        
		\xmark & \xmark & 3.41 & \hphantom{0}4.85 & 4.18 & \hphantom{0}7.64 \\
		\cmark & \xmark & 3.41 & \hphantom{0}4.81 & 4.12 & \hphantom{0}7.49 \\
		\xmark & \cmark & 6.38 & 10.22 & 8.26 & 14.86 \\
		\cmark & \cmark & 5.38 & \hphantom{0}8.80 & 6.56 & 12.79 \\[5pt]
   		\multicolumn{2}{@{}l}{\textit{Baseline:} MFCCs} & 10.95 & 13.55 & 20.94 & 29.4 \\
		\bottomrule
	\end{tabularx}
	\label{tbl:abx}
\end{table}

\textbf{Clustering metrics.}
To further analyze the discovered acoustic units, we compute four metrics used to evaluate clustering quality.
By mapping each unit to the overlapping phone label in the forced alignment, we evaluate the clustering quality in terms of the adjusted rand index (ARI), adjusted mutual information (AMI)~\cite{vinh+etal_icml09}, homogeneity, and completeness~\cite{rosenberg2007v}.
All these metrics are in the range $[0, 1]$, where higher is better.

Table~\ref{tbl:clustering-metrics} shows the results on the \texttt{dev-clean} subset of LibriSpeech.
Standardization gives consistent improvements across the metrics.
However, the clustering scores are relatively low overall.
This indicates that despite good phone discrimination scores, there is still a large gap between the discovered acoustic units and the ground-truth phonetic transcriptions.

\textbf{Number of clusters.} Next, we study the effect of the number of clusters on ABX score.
Table~\ref{tbl:clusters} reports ABX score (averaged over \texttt{dev-clean} and \texttt{dev-other}) for different numbers of clusters.
In contrast to the findings in~\cite{nguyen_zero_2020} (where $50$ clusters gave the best results), we
observe that increasing the number of clusters can improve ABX error rates.

\begin{table}[!t]
	\mytable
	\caption{Clustering metrics calculated on the $K$-means clustered CPC features, with and without prior standardization.}
	\eightpt % Interspeech
	\begin{tabularx}{\linewidth}{@{}ccccc@{}}
		\toprule
		Standardized & ARI & AMI & Homogeneity & Completeness \\
		\midrule
		\xmark & 0.221 & 0.450 & 0.477 & 0.425 \\
		\cmark & 0.255 & 0.488 & 0.517 & 0.462 \\
		\bottomrule
	\end{tabularx}
	\label{tbl:clustering-metrics}
	\vspace{-4mm}
\end{table}

\begin{table}[!b]
	\vspace{-3mm}
	\mytable
	\caption{ABX results for different numbers of K-means clusters.}
	\vspace{-1mm}
	\eightpt % Interspeech
	\begin{tabularx}{1.0\linewidth}{@{}Lcccc@{}}
	\toprule
	\# clusters & 50 & 100 & 150 & 200 \\
	\midrule
	Within & 7.09 & 6.73 & 6.74 & \textbf{6.68} \\
	Across & 9.68 & 9.15 & 9.20 & \textbf{9.09} \\
	\bottomrule
	\end{tabularx}
	\label{tbl:clusters}
\end{table}

\textbf{Feature selection.}
Finally, we investigate feature selection to improve acoustic unit discovery.
The idea is that speaker information might primarily be captured in a few specific dimensions of the CPC features.
To test this, we train a random forest to predict speaker labels for each frame of the CPC features.
We then prune the dimensions according to their importance ranking i.e., \ removing the dimensions that are most predictive of the speaker first.
Table~\ref{tbl:pruning} shows ABX results (averaged over \texttt{dev-clean} and \texttt{dev-other}) as a function of the number of retained dimensions.
We can see that ABX scores improve while pruning up to half of the feature dimensions.

\begin{table}[!b]
	\vspace{-2mm}
	\mytable
	\caption{ABX results after pruning the CPC dimensions that are least informative for predicting speaker.}
	\vspace{-1mm}
	\eightpt % Interspeech
	\begin{tabularx}{1.0\linewidth}{@{}Lccccccc@{}}
	\toprule
	\# features & 64 & 128 & 192 & 256 & 320 & 384 & 512 \\
	\midrule
	Within & \hphantom{0}8.88 & \hphantom{0}7.74 & 7.05 & 6.88 & \textbf{6.79} & 7.04 & 7.09 \\
	Across & 11.90 & 10.62 & 9.97 & 9.64 & \textbf{9.44} & 9.49 & 9.68 \\
	\bottomrule
	\end{tabularx}
	\label{tbl:pruning}
\end{table}

\subsection{Spoken language modeling}
\label{sec:lm}

\begin{table}[!t]	
    \mytable
    \caption{Results on the lexical, syntactic, and semantic spoken language modeling tasks.}
    \vspace{-1mm}
    \eightpt % Interspeech
    \begin{tabularx}{1.0\linewidth}{@{}Lcccc@{}}
        \toprule
        & & & \multicolumn{2}{c}{Semantic} \\
        \cmidrule(l){4-5}
        & Lexical & Syntactic & Synth. & Libri. \\
%        \cmidrule(l){3-4} \cmidrule(l){5-6} \cmidrule(l){7-8} \cmidrule(l){10-11}
        \midrule
        \multicolumn{2}{@{}l}{\textit{Topline:}} \\[2pt]
        Forced Align & 92 & 63 & 8.5 & 2.4 \\
        Phone & \textbf{98} & 67 & 12.2 & 20.2 \\ 
        RoBERTa & 96 & \textbf{82} & \textbf{33.2} & \textbf{27.8} \\%[5pt] 
        \midrule
        \multicolumn{2}{@{}l}{\textit{High budget:}} \\[2pt]        
        BERT baseline & 68 & 56 & 6.3 & 2.5 \\%[5pt] 
        \midrule
        \multicolumn{2}{@{}l}{\textit{Low budget:}} \\[2pt]                
        LSTM baseline & 61 & 53 & 7.4 & \textbf{2.4} \\
        LSTM speaker-norm & \textbf{65} & \textbf{54} & \textbf{9.2} & -1.1 \\
        Chorowski et al.~\cite{chorowski+etal_interspeech21} & 64 & 53 & 5.2 & -0.9 \\
        Maekaku et al.~\cite{maekak+etal_interspeech21} & 61 & \textbf{54} & 7.0 & -1.2 \\
        \bottomrule
    \end{tabularx}
    \label{tbl:language-modeling}
    \vspace{-4mm}
\end{table}

\textbf{Lexical: Spot-the-word.} To evaluate language models at the lexical level, we use the
spot-the-word task from \cite{le_godais_comparing_2017}.
In this task, models are presented with pairs comprising of an existing word and a similar non-word (e.g., ``brick'' and ``blick'').
The goal is to distinguish the word from the non-word by assigning it a higher probability.
An average classification accuracy is calculated over all word/non-word pairs.
Table~\ref{tbl:language-modeling} reports spot-the-word results 
on the sWUGGY~\cite{nguyen_zero_2020} test set. 
The set consists of 40k word/non-word pairs, generated using WUGGY~\cite{keuleers_wuggy_2010} and synthesized using Google Cloud Text-to-Speech.

\textbf{Syntactic: Acceptability judgments.}
At the syntactic level, we use grammar acceptability judgments to test the language models.
This is similar to the spot-the-word task, but the goal is to distinguish grammatical from ungrammatical sentences (for example, ``the dogs \underline{eat} meat'' versus ``the dogs \underline{eats} meat'').
Table~\ref{tbl:language-modeling} reports classification accuracy on sBLIMP~\cite{nguyen_zero_2020}, a spoken version of the BLIMP \cite{warstadt_blimp_2020} benchmark.
The sBLIMP test set consists of 64k sentence pairs covering 12 grammar categories, e.g.,\ anaphor agreement, island effects, and subject-verb agreement.

\textbf{Semantic: Similarity judgments.}
We use human similarity judgments between word pairs to assess the semantic information captured by the language models.
First, human evaluators score pairs of words (e.g.,\ ``abduct'' and ``kidnap'') based on their semantic similarity.
Next, we extract a fixed-dimensional representation for each word by pooling the outputs of a hidden layer of the language model.
Specifically, we follow~\cite{nguyen_zero_2020} by applying min-pooling to the outputs of the second LSTM layer.
Finally, we compute the cosine similarity between the two representations and evaluate how well it compares to the human similarity scores with results reported as the Spearman’s rank correlation coefficient.
Table~\ref{tbl:language-modeling} reports semantic similarity scores on the sSIMI benchmark \cite{nguyen_zero_2020}, a combination of 13 existing semantic similarity and relatedness tests including both synthetic and natural speech.

\textbf{Results summary.}
In the bottom section of Table~\ref{tbl:language-modeling}, we compare our approach (LSTM speaker-norm) to the three low-budget
models submitted to the ZeroSpeech2021 challenge~\cite{nguyen_zero_2020, chorowski+etal_interspeech21, maekak+etal_interspeech21}.
In the low-budget category, the LSTM language model trained on clustered speaker normalized CPC features scores the best on the lexical and syntactic tasks.
This shows that better ABX scores (through speaker normalization) translate into better results on spoken language modeling.
However, there remains a large gap in performance compared to the supervised topline systems.
While the toplines score well on the lexical task, the syntactic and semantic results show that there is still room for improvement (despite access to ground-truth transcriptions).
This suggests that syllable- or word-like units may be required for these tasks.
Finally, while our approach doesn't match the performance of the high-budget BERT baseline, we expect the speaker normalization step to benefit this model as well.

\section{Conclusion}

We proposed a simple speaker normalization method for contrastive predictive coding (CPC) models.
By analyzing a CPC	model, we found that speaker information is largely captured by the per-utterance mean of the features.
Based on this observation, we showed that standardizing the features effectively removes speaker details.
We incorporated this speaker normalization step into systems for acoustic unit discovery and spoken language modeling, improving the ZeroSpeech 2021 Challenge baselines.

\vspace{2pt}
{\eightpt
\noindent \textbf{Acknowledgements.}
This work is supported in part by the South African NRF (grant no.\ 120409), a Google PhD Scholarship for BvN, a DeepMind Scholarship for LN, and a Google Faculty Award for HK.}

\bibliography{mybib}

\end{document}